\documentclass[aps,prl,reprint]{revtex4-1} 
\pdfoutput=1
\usepackage{graphicx} 

\begin{document} 
 
\title{Answer to the Comment of John Arrington.} 
  
\author{J.\,C. Bernauer et al., A1 Collaboration at MAMI}
\affiliation{Institut f\"ur Kernphysik, Johannes  
  Gutenberg-Universit\"at Mainz, 55099 Mainz, Germany.}
  
\date{\today}  
 
\maketitle  
 
The criticism of our results is based on the ``modern and complete calculations'' 
of the Coulomb correction of Sick and Trautmann \cite {Sick:1998} going back to work 
of Lewis published in 1956 \cite{Lewis:1956zz}. This calculation is done 
in ``second Born approximation'', i.e. TPE without intermediate excited states. 
The integral describing the TPE has been evaluated numerically \cite{Drell:1962} and 
it is apparently this code on which Arrington's Coulomb corrections are based. 

\begin{figure}[h] 
\includegraphics[]{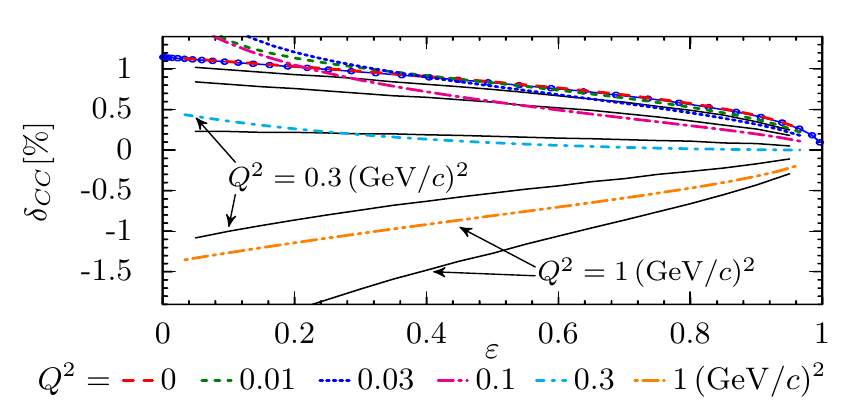} 
\caption{\label{fig:1} Comparison of Coulomb corrections.} 
\end{figure}

In order to quantify the influence of TPE on our results we have chosen the modern 
analytical integration by Borisyuk and Kobushkin (ref.\,[4] of the Comment) lending 
itself to an easy calculation. Figure~\ref{fig:1} shows Fig.~1 of the Comment overlayed 
with these calculations for the same $Q^2$. It demonstrates the variance of TPE 
calculations also indicated by a remark in the caption of this figure in the Comment. 
All calculations go to the same curve in the limit $Q^2 \rightarrow 0$ given by 
eq.\,(1) of the Comment.
\vspace*{-2mm}   

The uncertainty is also demonstrated in Fig.~2 of reference \cite{Meziane:2010xc} 
showing five theoretical calculations of TPE, which are mutually inconsistent, with 
all but one disagreeing with the null experimental TPE effect at $Q^2 =2.5$\,(GeV/c)$^2$. 
Though that work concerns polarization variables and not a Rosenbluth separation, the 
diagrams of TPE are based on QED and have to be valid for both. 

All this is not surprising since the unconstrained part of the TPE amplitude 
resulting from the off-shell internal structure of the nucleon does cause considerable 
variance at present among the different TPE calculations. Such calculations require 
knowledge beyond on-shell form factors, and imply as well dispersion effects resulting 
from the excitation spectrum of the nucleon.

\begin{figure}[h]
\includegraphics[width=7.5cm]{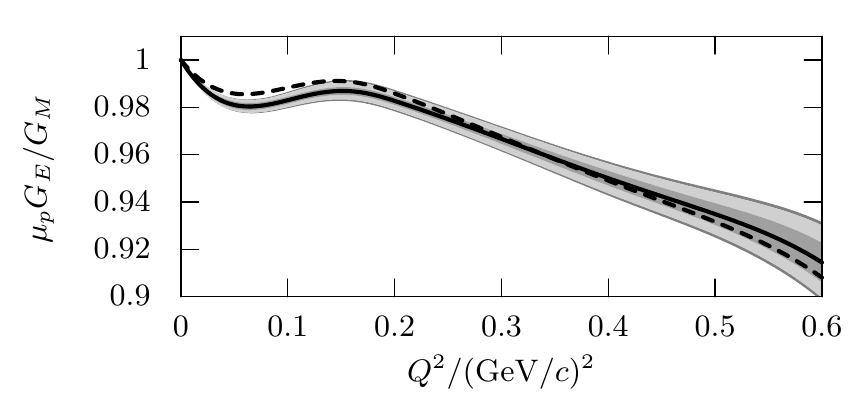} 
\caption{\label{fig:2} $\mu_p G_E/G_M$ with Coulomb correction as
  published (solid), with TPE acc.\ to Borisyuk and Kobushkin (dashed).}
\end{figure}

Nevertheless, we have applied the calculation of Borisyuk and Kobushkin to our data 
and refitted. Figure~\ref{fig:2} shows the ratio of the electric over the magnetic 
form factors $\mu_p G_E/G_M$ with the spline ansatz. 
We also determined the radii with all models for $G_E$ and $G_M$ as in our Letter. 
The averaged result changes $\left\langle  r^2\right\rangle^\frac{1}{2}$ without 
$\rightarrow$ with TPE correction:
\begin{eqnarray} 
\left\langle r_{E}^2\right\rangle^\frac{1}{2}&=&0.879 (8) \rightarrow 0.876 (8)\, \mathrm{fm},\nonumber\\  
\left\langle r_{M}^2\right\rangle^\frac{1}{2}&=&0.777 (17) \rightarrow 0.803 (17)\,\mathrm{fm}.\nonumber 
\end{eqnarray} 
We wrote in our Letter: {\it These radii have to be taken with the applied corrections 
in mind. While the effect of the Coulomb correction used is compatible with other 
studies  {\rm (see references in our Letter)} a more sophisticated theoretical calculation may 
affect the results slightly.}

Finally, the statements about our systematic errors are wrong. The statistical contributions 
to the point-to-point systematic errors are shown to be Gaussian and are therefore taken 
together with the statistical counting error (innermost error band in our Letter). The 
systematic uncertainties due to the angular dependences are linearly added to this statistical 
error and shown by the second band. For the outermost band the Coulomb correction has been 
varied by $\pm 50\%$. For details see ref.\,[8] of the Comment.

In summary, the criticism of the Comment neglects the uncertainty of the TPE corrections and 
exaggerates the quantitative effect at small $Q^2$. We hold that we are well advised to apply 
only the Coulomb correction of the unique limit at $Q^2 = 0$. 
In the detailed follow-up paper we intend to present the experimental effect of TPE in a way 
making a comparison to theoretical calculations possible without reanalysis of the data.   
 
\acknowledgments{We are indebted to Marc Vanderhaeghen for advising 
us on TPE corrections.}


\end{document}